\documentclass{elsart}

\usepackage{graphicx}
\usepackage{multicol}

\begin{document}
\begin{frontmatter}



\title{Evaporation residues produced in spallation of $^{208}$Pb by protons at 500A~MeV}


\author[Orsay]{L.~Audouin\corauthref{thesis}}, \ead{audouin@ipno.in2p3.fr}
\author[Orsay]{L.~Tassan-Got},
\author[GSI]{P.~Armbruster},
\author[Santiago]{J.~Benlliure},
\author[Orsay]{M.~Bernas},
\author[DAPNIA]{A.~Boudard},
\author[Santiago]{E.~Casarejos},
\author[CENBG]{S.~Czajkowski},
\author[GSI]{T.~Enqvist},
\author[DAPNIA]{B.~Fern\'andez-Dom\'inguez},
\author[GSI]{B.~Jurado},
\author[DAPNIA]{R.~Legrain\thanksref{dead}},
\author[DAPNIA]{S.~Leray},
\author[Orsay]{B.~Mustapha},
\author[Santiago]{J.~Pereira},
\author[CENBG]{M.~Pravikoff},
\author[Orsay]{F.~Rejmund},
\author[GSI]{M.-V.~Ricciardi},
\author[GSI]{K.-H.~Schmidt},
\author[Orsay]{C.~St\'ephan},
\author[Orsay,GSI]{J.~Taieb},
\author[DAPNIA]{C.~Volant},
\author[DAPNIA]{W.~Wlaz\l o}

\corauth[thesis]{This work forms part of the PhD thesis of L. Audouin}
\thanks[dead]{Deceased}

\address[Orsay]{IPN Orsay, CNRS-IN2P3, Campus Paris XI, 91406 Orsay, France}
\address[GSI]{GSI, Planckstrasse 1, 64291 Darmstadt, Germany}
\address[Santiago]{Universitad Santiago de Compostela, 15706 Santiago de Compostela, Spain}
\address[DAPNIA]{DAPNIA/SPhN, CEA/Saclay, F-91191 Gif-sur-Yvette cedex, France}
\address[CENBG]{CENBG, CNRS-IN2P3, 33175 Gradignan, France}

\begin{abstract}
The production cross sections of fragmentation-evaporation residues in the reaction Pb+p at 500A~MeV have been measured using the inverse-kinematics method and the FRS spectrometer (GSI). Fragments were identified in nuclear charge using ionisation chambers. The mass identification was performed event-by-event using the $B\rho-TOF-\Delta E$ technique. Although partially-unresolved ionic charge states induced an ambiguity on the mass of some heavy fragments, production rates could be obtained with a high accuracy by systematically accounting for the polluting ionic charge states. The contribution of multiple reactions in the target was subtracted using a new, partly self-consistent code. The isobaric distributions are found to have a shape very close to the one observed in experiments at higher energy. Kinematic properties of the fragments were also measured. The total and the isotopic cross sections, including charge-pickup cross sections, are in good agreement with previous measurements and models. The data are discussed in the light of previous spallation measurements, especially on lead at 1~GeV.
\end{abstract}

\end{frontmatter}

\newcommand{\stwo}{$S_2$}
\newcommand{\sfour}{$S_4$}
\newcommand{\mg}{~mg/cm$^2$}
\renewcommand{\etal}{{\em et~al.}}


\section{Introduction}


Spallation reactions, i.e. proton-induced reactions on heavy targets at a few hundred MeV, have been the subject of many studies since 1950. They are known to be a valuable tool for the study of the de-excitation of hot nuclei because, contrarily to reactions between heavy ions, they lead to the formation of hot prefragments with only a limited excitation of the collective degrees of freedom such as rotation or compression. Their study has also been motivated by astrophysics, as cosmic rays undergo spallation reactions with the hydrogen and helium nuclei of the interstellar medium.

Recently, progresses in high-power accelerator technologies have made possible the realisation of intense neutron sources based on spallation reactions. Such sources are needed for Accelerator-Driven Systems~\cite{Bowmann,Rubbia}, and also find applications in nuclear physics~\cite{NToF} and for material physics and biology~\cite{ESS}. Furthermore, spallation reactions can also be used to produce exotic nuclei and, hence, secondary beams~\cite{ISOLDE,EURISOL}. Those new applications have motivated a large number of experimental works and created a strong demand for high-precision calculation codes.

In order to bring new data and therefore new constraints for the codes, measurements of reaction residues have been undertaken at GSI by an international collaboration. These experiments are based on the inverse-kinematics method. Fragments are identified in-flight using the FRS spectrometer~\cite{FRS}, making possible the first measurements of complete nuclide distributions. In the frame of those studies, production cross sections have already been published for several systems: Au+p at 800A~MeV~\cite{Au_Fanny,Au_Pepe}, Pb+p at 1A~GeV~\cite{Wlazlo,Pb_p}, Pb+d at 1A~GeV~\cite{Pb_d}, U+p at 1A~GeV~\cite{Taieb,Monique} and U+d at 1A~GeV\cite{U_Enrique,U_Jorge}.

These results have helped to partially discriminate between the respective influence of the two main steps of the spallation process, the intranuclear cascade and the fission/evaporation process. The behavior of several codes (the ISABEL~\cite{ISABEL}, INCL4~\cite{INCL4} or BRIC~\cite{BRIC} intranuclear cascades, the ABLA fission/evaporation code~\cite{ABLA}) has proved to be now overall satisfactory for proton energies around 1 GeV. On the other hand, important failures in the description of the emission of charged particles in the Dresner evaporation code~\cite{Dresner} have been put in evidence~\cite{Au_Fanny}. In the 1 GeV energy region, the only serious, remaining deficiency is the underestimation of the lightest evaporation products, which are related to the most violent collisions. Despite large differences in the description of the spallation process, all the codes mentioned above present this weakness. This indicates that some phenomena have not been taken into account. In recent experiments conducted at GSI on lighter nuclei ($^{56}$Fe, $^{136}$Xe), our collaboration explored a range of nuclear temperatures higher than in the systems mentioned in the previous paragraph. Indications were found that fast break-up decay may play an important role in high-energy spallation reactions~\cite{Paolo}.

The question of understanding of the evolution of the reaction mechanisms with decreasing energy also remains open. This is an important point in the perspective of technical applications, because nuclear reactions in thick targets happen in a broad energy range: beam particles are subject to electronic slowing down, and also fast particles emitted in the first stage of the reactions are likely to produce additional nuclear reactions, giving rise to an {\it internuclear} cascade. To address the question of the dependance of the reaction on the energy of the incident particle, an experiment has been performed at GSI aiming at measuring production cross sections of residues in the spallation of lead by protons at 500A~MeV.

The present paper deals with the experimental results on the fragmentation-evaporation residues obtained in this experiment. It completes  the results already published obtained during the same experiment for the fission products~\cite{NPA_Bea}. Detailed confrontations between the results from codes dedicated to the description of the spallation process and these data as well as other data on evaporation residues obtained by our group and other related measurements on spallation reactions like light particle production are postponed to a forthcoming paper.

The energy chosen for this experiment, which is low in comparison to the FRS standards for experiments involving nuclei as heavy as lead~\cite{achromat}, was a source of difficulties for the identification of the fragments in the spectrometer. The modified setting and the analysis methods developed for this experiment have been presented in a dedicated paper~\cite{NIM_loa}. We will briefly recall their main features in section~\ref{chap:setup}. We will then discuss in section~\ref{chap:secondary} the influence of multiple reactions taking place in the liquid hydrogen target and modifying the observed fragment production, and the method which has been employed to remove their contribution. In section~\ref{chap:kinematics} we will present the results on the reaction kinematics. Finally, in section~\ref{chap:results} we will present the production cross sections.


\section{Experimental setup and analysis process} \label{chap:setup}


The GSI synchrotron (SIS) was used to produce a 500A~MeV $^{208}$Pb pulsed beam with a pulse duration of 4 seconds and a repetition time of 8 seconds. The beam was sent onto a 87.3\mg~liquid hydrogen target~\cite{target} located at the entrance of the FRagment Separator (FRS). The target window consisted of two 9\mg~Ti foils on each side. The beam intensity was monitored during all the experiment by a beam-intensity monitor~\cite{SEETRAM}. In order to maximise the proportion of fully stripped fragments in the spectrometer, a 60\mg~Nb foil was placed after the target.

\begin{figure}[ht]
\begin{center}
\includegraphics[width=0.9\textwidth]{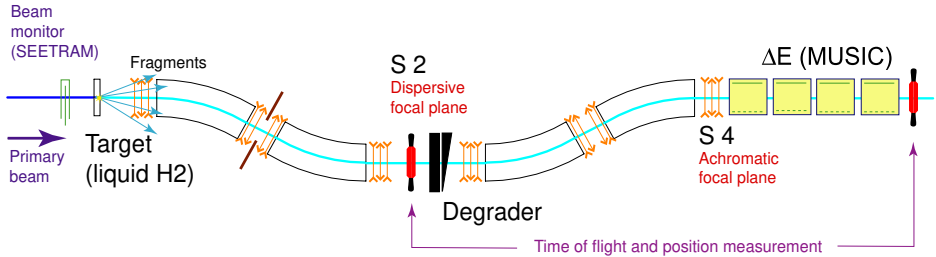}
\caption{Schematic view of the FRagment Separator. Each magnetic section between focal planes (the target location, $S_2$ and $S_4$) consists of two dipoles plus several quadrupoles and sextupoles (the latter are not represented here as they were not used during this experiment).}
\label{fig:FRS}
\end{center}
\end{figure}

Fragments were identified in-flight using the FRS spectrometer (see figure~\ref{fig:FRS}). The rigidity of the fragments in each of the two magnetic sections is given by:

\begin{equation}
B \rho = \frac{m_0 c}{e} \frac{A}{q} ( \beta \gamma )
\label{eqn:brho}
\end{equation}

where $B$ is the magnetic field, $\rho$ the curvature radius of the fragment trajectory, $A$ the mass number, $q$ the ionic charge of the fragment, and $\beta$ and $\gamma$ are the Lorentz relativistic coefficients. The presence of $q$ in equation~\ref{eqn:brho} is a critical point, as a large part of the fragments produced at the energy of 500A~MeV chosen for this experiment were not fully stripped.

In order to measure the nuclear charge ($Z$) of the fragment, 4 MUlti-Sampling Ionisation Chambers (MUSIC; see~\cite{MUSIC} for complete description) were placed at the exit of the spectrometer, each one filled with 2 bar of P10 gas mixture (90\% Ar, 10\% CH$_4$). The total gas thickness was 800\mg, with the measurement of the energy loss effectively performed only in some 2/3 of the length of each chamber. Using such a large gas thickness was necessary in order to maximise the charge exchanges (electron pick-up and stripping) of the fragment in the gas, thus washing out the influence of the incoming ionic charge state of the fragment on the energy loss, and ensuring that the latter represents the nuclear charge of the fragment with sufficient resolution~\cite{NIM_loa}. The obtained resolution,  $\Delta$Z{\it (FWHM)}/Z, ranged between 0.6\% for the lightest fragments and 0.9\% for the heaviest.

The horizontal position of the fragments at the intermediate and final focal planes (respectively $S_2$~and $S_4$; see figure~\ref{fig:FRS}) was measured using 3~mm thick plastic scintillators. The signals of these detectors were also used to measure the time of flight of the fragments in the second part of the spectrometer. The $A/q$~ratio of the fragment was deduced from the combination of these measurements, according to equation~\ref{eqn:brho}.

The thick aluminium degrader (1700\mg) located at the intermediate focal plane (\stwo) was used as a passive energy-loss measurement device. As the energy loss can be related to the variation of the magnetic rigidity and the ionic charge states in the two magnetic sections, the latter may thus be deduced from the energy loss as obtained from the nuclear-charge identification and the velocity measurement~\cite{NIM_loa}.

The resolution obtained in this measurement was not sufficient to discriminate the ionic charge states on an event-by-event basis. The only information obtained was the charge-state changing between the first and the second part of the spectrometer, an integer value that we will note $\Delta q$. Besides, the evaluation of the variation of magnetic rigidity was also used to reject fragments that underwent a nuclear reaction at \stwo.

The mass of each fragment was determined assuming that its number of electrons in the FRS was the minimum required by its $\Delta q$ value~\cite{NIM_loa}. The production rate for each nuclide was then calculated by constructing its full velocity distribution in the first part of the FRS, using formula~\ref{eqn:brho}. For many nuclides, the momentum width was larger than the momentum acceptance of the FRS ($\pm1.5\%$); in this case several settings of the magnets were used in order to cover the full momentum distribution of the fragment. Due to the hypothesis made on the ionic charge state, some fragments were misidentified; the corresponding correction factor for the production rates was deduced from ionic charge-state probabilities calculated using the code GLOBAL~\cite{GLOBAL}.

The above procedure was performed separately for each group of fragments characterised by a given $\Delta q$ value. The probability of each $\Delta q$ value was then deduced from the scaling factor necessary for all isotopic distributions obtained for a given element to match with each other. The obtained values were found to be in good agreement with the GLOBAL calculations (discrepancies lower than 10\% for the most abundant ionic charge-state combinations, and less than 20\% for other combinations)~\cite{NIM_loa}.

The production rates were corrected for the losses in the different layers of matter located in the path of the fragments after the target area (degrader, plastic scintillators, MUSIC chambers). The total reaction cross sections were calculated using the Karol optical-model-based code~\cite{Karol}. Losses were found to be of the order of 30\%, including reactions in the MUSIC chambers (the latter being characterised by signals of first and last chambers being improperly correlated). The reaction rates in the different layers of matter are presented in table~\ref{table:reac_prob}. The dead time of the acquisition and the detector efficiencies were also taken into account. Fragment losses due to limited angular acceptance of the FRS were found to be negligible. All the measurements and the analysis procedure above were repeated with an empty target, and the resulting production rates were subtracted from the total production rates. Finally, the production cross sections were obtained by normalising the production rates to the number of atoms in the liquid hydrogen target, which had been measured in a previous experiment, and to the beam intensity.

\renewcommand{\arraystretch}{0.4}
\begin{table}[ht]
\begin{center}
{\footnotesize
\begin{tabular}{|c|c|c|c|c|c|c|}
\hline
& & & & & \\
Layer                & Target windows & Stripper & Scintillator & Degrader  & MUSICs \\
& & & & & \\

\hline

                     & \multicolumn{2}{c|}{}     & \multicolumn{2}{c|}{}    & \\
Focal plane          & \multicolumn{2}{c|}{$S_0$}&\multicolumn{2}{c|}{$S_2$}& $S_4$  \\
                     & \multicolumn{2}{c|}{}     & \multicolumn{2}{c|}{}    & \\

\hline

& & & & & \\
Reaction             & 2.1\%          & 2.1\%    & 8.4\%        & 16.6\%    & 4.9\% \\
probability          & & & & & \\
& & & & & \\

\hline
                     & \multicolumn{2}{c|}{}     & \multicolumn{2}{c|}{}    & \\
Method of            & \multicolumn{2}{c|}{{\it none}} & \multicolumn{2}{c|}{$B\rho$ change} & $\Delta E$ change\\
rejection            & \multicolumn{2}{c|}{}     & \multicolumn{2}{c|}{}    & \\
                     & \multicolumn{2}{c|}{}     & \multicolumn{2}{c|}{}    & \\
\hline

                     & \multicolumn{2}{c|}{}     & \multicolumn{3}{c|}{} \\
Method of            & \multicolumn{2}{c|}{Dedicated measurement} & \multicolumn{3}{c|}{Calculation}    \\
correction           & \multicolumn{2}{c|}{(empty target)}        & \multicolumn{3}{c|}{(Karol model)}    \\
                     & \multicolumn{2}{c|}{}     & \multicolumn{3}{c|}{} \\

\hline
\end{tabular}
}
\caption{Reaction probabilities of the beam ($^{208}Pb$ at $500A~MeV$ at the entrance of the FRS) in the various layers of matter of the FRS beam line. Rejection method of the formed nuclei and correction method of the production rates are also mentioned. See text for details.}
\label{table:reac_prob}
\end{center}
\end{table}
\renewcommand{\arraystretch}{1}


\section{Secondary reactions in the target} \label{chap:secondary}


Any fragment formed in a collision with a proton of the target may undergo additional nuclear collisions, in the target as well as in surrounding material (target window, stripper foil). These secondary reactions are expected to play an important role in the production of nuclides far from the projectile: at relativistic energies, proton-induced reactions mainly produce nuclei lighter than the heavy partner of the reaction, therefore in most cases multiple reactions will remove more nucleons than a single reaction.

There is no way to identify such multiple reactions during the analysis process. Therefore, one has to unfold their contribution by using calculated reaction cross sections or by performing a self-consistent calculation. This section is dedicated to the presentation of a new method developed for this experiment aiming at estimating the contribution of the multiple reactions with a high precision while minimising the input from codes.

\subsection{Unfolding method}

The production cross section of a nuclide $f$ from projectile (indexed as $0$ in the following) is written as:

\begin{equation}
\sigma_{0\rightarrow f} = \frac{e^{\frac{\sigma_0+\sigma_f}{2}x}}{x} \;
\left(T_f(x) - \frac{x^2}{2} \sum_{A_0 < A_i < A_f}^{} \;
\sigma_{0\rightarrow i} \; \sigma_{i\rightarrow f} \;
e^{-\frac{\sigma_0+\sigma_i+\sigma_f}{3}x} \right)
\label{eqn:secondary}
\end{equation}

where $\sigma_i$ is the total reaction cross sections of a nuclide $i$ on a nuclide of the target, $\sigma(i,j)$ is the production cross section of a  nuclide $(A_j,Z_j)$ from a nuclide $(A_i,Z_i)$, $T$ is the observed production rate, and $x$ is the thickness of the target. A derivation of this equation is presented in appendix~\ref{chap:sec_calc}. This corresponds to a first-order approximation (i.e. only double reactions in the hydrogen are taken into account), but it is easily extended to higher orders. In our calculations, we actually accounted for the second order reactions: triple reactions in hydrogen, and reactions involving one reaction in a target window and one in the hydrogen (or the reverse). We found that those second order terms actually accounted for less than 20\% of the multiple reactions.

Solving this system of equations (one equation for each observed nuclide) requires the calculation of all the $\sigma$ terms. For the total cross sections, several reliable codes exist; we used the optical-model-based code of Karol~\cite{Karol}, the same one we used for the probability of nuclear reactions at the intermediate focal plane of the FRS. For the partial cross sections involving heavy target nuclei (target windows and stripper foil), the EPAX parametrisation~\cite{EPAX} offers reliable results with minimal calculation time. In the case of proton-induced reactions, the Monte-Carlo cascade-evaporation codes would seem an obvious choice, but they could hardly be used here for two reasons. First, the calculation required to evaluate the hundreds of possible reactions would have been very time consuming. Second, as one of the goals of this experiment was to produce data to constrain these codes in this poorly-known energy region, the use of those codes might have introduced an artificial consistency between the data and the codes.

In order to calculate the isotopic cross sections, we can decompose each cross section of proton-induced reactions in a product of 3 factors:

\begin{equation}
\sigma(x,y) = \sigma_x \; P_A((A_x,Z_x) \rightarrow A_y )
\; P_Z((A_x,Z_x,A_y)\rightarrow Z_y)
\end{equation}

Here, the first term is the total reaction cross section of a nuclide $x$ (we have already stated that it could be calculated using the Karol formula~\cite{Karol}), the second term is the probability to form a nuclide of mass $A_y$ from a nuclide of mass $A_x$, and the third term is the probability that the nuclide formed with a mass $A_y$ has an atomic number $Z_y$.

In order to estimate the second term, we used a property of proton-induced spallation reactions: nearly all nuclides $b$ formed from a nuclide $a$ have a mass strictly smaller than the one from $a$. For example, in the 500A~MeV experiment on $^{208}$Pb we observed no formation of any nuclide of mass 209, and nuclides of mass 208 ($^{208}$Bi) are formed in less than 0.1\% of the reactions. Furthermore, we assumed that, as far as only the probability of mass loss is concerned, the influence of the isospin of the target nuclide is weak enough to be neglected. Using these assumptions, one can solve the system of equations~\ref{eqn:secondary} isobar by isobar, in the decreasing masses order, because the term $P_A(A_x \rightarrow A_y )$ required by each equation is immediately obtained from the previously corrected data as $P_A(A_0-(A_x-A_y))$ (where $A_0$ is the projectile mass).

For the third term, this kind of simple scaling law cannot be applied because, although $P_Z$ depends only on $A_y$ for large values of $A_x-A_y$. Indeed, this well-known property defines the so-called residue corridor~\cite{Dufour}: the statistical nature of the evaporation process favors its ending close to nuclei for which the probability to evaporate a neutron and a proton (in other words, the neutron and proton separation energies), are the closest, a property which is completely independent of the entrance channel. But, in the case of short evaporation chains, the influence of the entrance channel is not suppressed; in other words, for low values of $A_x-A_y$, $P_Z$ depends not only on $A_y$ but also on $Z_x$ . This memory effect is fully taken into account in the EPAX parametrisation~\cite{EPAX}. We will describe hereafter how the EPAX formula can be used even though it is out of its energy domain applicability. Please note that, as EPAX does not take into account the fission process, this method would not be appropriate for reactions involving highly fissile nuclei such as uranium.

\subsection{Calculation of isobaric distributions using EPAX}

Some characteristics of the EPAX parametrisation~\cite{EPAX} correspond to the requirements for a multiple reactions calculation: it needs very little computing time, and it proved to be reliable, not only for reactions involving nuclides close to the stability valley, but also for proton-rich nuclides~\cite{112Sn}.

\begin{figure}[ht]
\begin{center}
\includegraphics[width=0.9\textwidth]{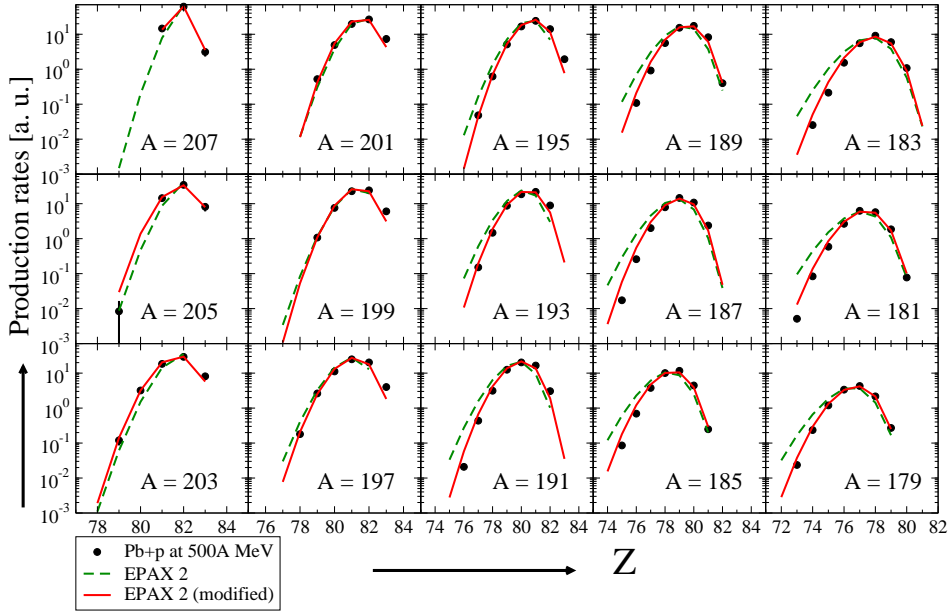}
\caption{Comparison of production rates obtained in the Pb+p at 500A~MeV experiment (dots) with calculations performed with both the standard (discontinuous lines) and a version we modified (continuous lines) of the EPAX parametrisation~\cite{EPAX}. For each isobaric spectrum, calculations have been renormalized to the data.}
\label{fig:EPAX}
\end{center}
\end{figure}

EPAX has been written in order to reproduce residues from reactions in the limiting-fragmentation regime~\cite{limit_frag}, which is reached in spallation  only for projectile energies of several GeV. The mass distribution of residues formed in 1A~GeV proton-induced spallation reactions exhibits a very different shape from the one formed in the limiting-fragmentation regime~\cite{Pb_p}. Therefore, the residues from the same reaction with half the incident energy may certainly not be reproduced by the mass-loss formula of EPAX.

On the other hand, the shape of the isobaric spectra is mainly a consequence of the sequential evaporation mechanism. Therefore, there is no reason why its validity should be limited to high incident energies. To check this assumption we extracted the isobaric component of the EPAX formula and compared it to our data after proper renormalization for each isobaric spectrum. Only minor adjustments were necessary to obtain a very satisfactory reproduction of the measured production rates, as it can be seen in figure~\ref{fig:EPAX}. Such a comparison makes sense as the secondary reactions are not expected to play an important role in the production of nuclides with a mass loss of less than, roughly, 30 mass units with respect to the projectile. Unexpectedly, we observed that the charge-pickup reactions were also reasonably well reproduced by the parametrisation, despite the fact that the EPAX authors didn't take this phenomena into account during the development process.

\subsection{Contribution of the multiple reactions in the target}

\begin{figure}[ht]
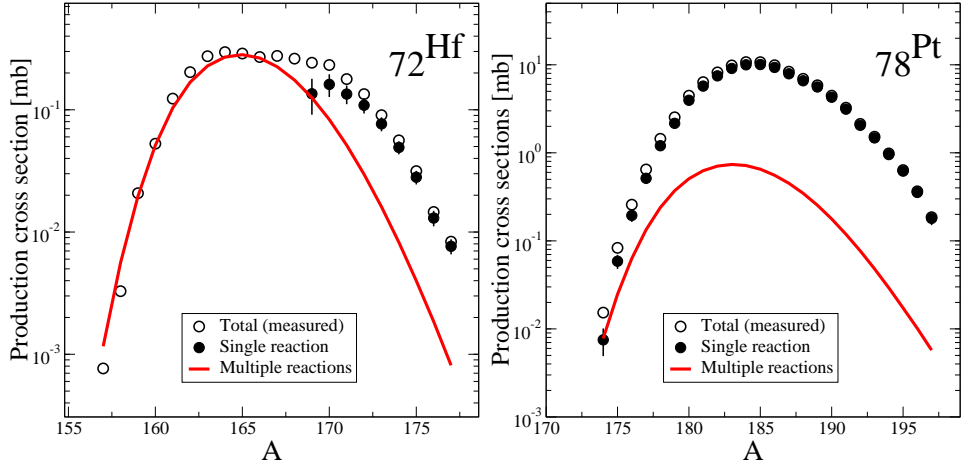

\begin{center}
\includegraphics[width=0.45\textwidth]{secondary72.eps}
\includegraphics[width=0.45\textwidth]{secondary78.eps}
\caption{Production cross sections in hydrogen before and after the subtraction of the multiple reactions (empty and full dots, respectively), and contribution of the multiple reactions in the target (continuous lines), for Hf (left) and Pt (right) isotopes.}
\label{fig:secondary}
\end{center}
\end{figure}

The results of multiple-reaction calculations are presented in figure~\ref{fig:secondary} for 2 isotopic distributions, each one corresponding to an extreme situation regarding the contribution of the multiple reactions. For $Z$ around 78, the multiple reactions are an important contributor for very proton-rich nuclides only. Their contribution increases and spreads towards neutron-rich nuclides with decreasing $Z$. The very proton-rich part of the isotopic distributions of the light elements such as Hf is reproduced by our calculation with differences being less than 20\% in most cases. This demonstrates the validity of our approach. As the uncertainty on this calculation could not be estimated in a systematic way, we quoted the value of 20\% mentioned above.

We chose to consider as results of the experiment only the cross sections deduced from production rates for which multiple reactions contributed for less than 50\%. This discards nearly all nuclides with $Z<70$, which represent only a very small fraction of the fragmentation residues.


\section{Kinematics of the reaction} \label{chap:kinematics}


Once nuclei are identified, their velocity in the first part of the FRS can be calculated using the equation~\ref{eqn:brho}:

\begin{equation}
( \beta \gamma )_1 = \frac{(B\rho)_1}{A/q_1}
\end{equation}

Here the index 1 stands for the first part of the FRS (before \stwo). Using this technique, the resolution is expected to be of the same order as the one obtained for the magnetic rigidity, roughly $5.10^{-4}$. At the energy used in this experiment, this is by a factor of 3 better than what can be achieved by a time-of-flight measurement in the second part of the FRS. This high resolution makes the FRS a remarkable tool to study the kinematics of nuclear reactions.

We have already pointed out that, because of the limited momentum acceptance of the FRS, the reconstruction of the full velocity spectra of each nuclide is a necessary step in order to evaluate the production rates (section~\ref{chap:setup}). The measured velocity spectra are Lorentz transformed into the reference frame of the beam, and corrected for the contribution of the beam width and for the velocity scattering due to the passage through the target and the surrounding materials. The resulting spectra give direct access to the longitudinal momentum transfer and to the longitudinal momentum spread caused by the nuclear reactions.

\begin{figure}[t]
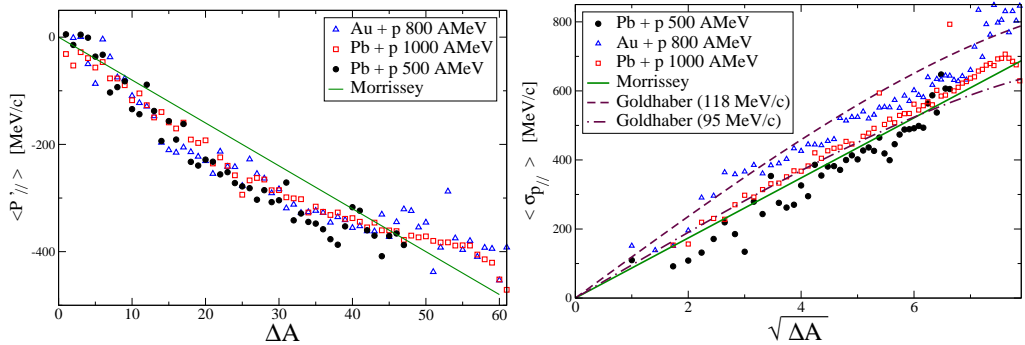

\begin{center}
\includegraphics[width=0.48\textwidth]{p_parrallel_mean.eps}
\includegraphics[width=0.48\textwidth]{sigma_mean.eps}
\caption{Momentum transfer (left) and momentum width (right) measured in the reactions Pb+p at 500A~MeV (triangles), Pb+p at 1A~GeV (squares) and Au+p at 800A~MeV (circles). Data are compared to Morrissey systematics~\cite{Morrissey} (continuous lines) and Goldhaber formula~\cite{Goldhaber}, the later being computed with a Fermi momentum of 118~MeV.c$^{-1}$ (dashed line) and 95~MeV.c$^{-1}$ (dashed-dotted line). All data have been normalised following the Morrissey prescription.}
\label{fig:kin}
\end{center}
\end{figure}

In figure~\ref{fig:kin} these quantities are compared to results of previous spallation experiments as well as to the well-known Morrissey systematics~\cite{Morrissey} and to the Goldhaber formula~\cite{Goldhaber}. The data were averaged over each isobaric distribution using the production cross sections as weighting factor.

A very similar tendency is obtained for all the experimental data regarding the momentum transfer. The simple linear dependence proposed by the Morrissey systematics is not fulfilled by the experiment. The longitudinal momentum transfer is underestimated for fragments corresponding to a mass loss of 10 to 45 units with respect to the projectile. This underestimation vanishes with increasing mass losses.

The momentum width exhibit a linear dependance to the square root of the mass loss. The Morrissey systematics offers a fair reproduction of the data. Using the result of a direct measurement of the Fermi momentum (118~MeV/c)~\cite{fermi_mes}, the Goldhaber formula overestimates the momentum width. This is not unexpected as this formula only takes into account the nucleons removed during the cascade stage, which lead to larger momentum fluctuations with respect to the nucleons emitted in the evaporation phase~\cite{Hanelt}. Nevertheless, a better agreement with data can be obtained by using an arbitrary Fermi momentum value of 95~MeV/c, as often done in heavy-ion calculations. The dispersion between the different data sets is probably related to the delicate corrections applied to the data, namely the beam width in momentum and position, which are difficult to estimate.


\section{Production cross sections} \label{chap:results}


\subsection{Isotopic cross sections}

\begin{figure}[ht]
\begin{center}
\includegraphics[width=0.9\textwidth]{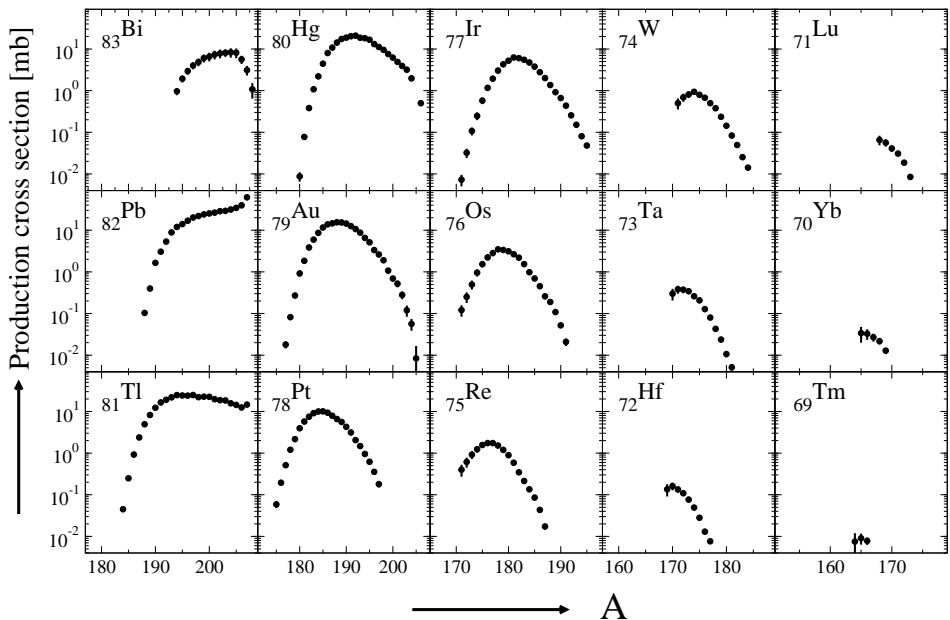}
\caption{Isotopic production cross-section distributions of residues in the reaction Pb+p at 500A~MeV.}
\label{fig:cs}
\end{center}
\end{figure}

Figure~\ref{fig:cs} shows the measured distributions of isotopic production cross-sections for elements between erbium and bismuth (see appendix~\ref{chap:annexe_xs} for the full list of cross sections). Some 250 spallation-evaporation cross sections have been measured. One observes that the cross sections of isotopic chains vary smoothly. As no even-odd fluctuations are expected~\cite{Valentina}, and considering the statistical nature of the evaporation process, this corroborates that the measured production rates do not hide any other source of fluctuations. The upper limit of 50\% production of multiple reactions in the target, that we have decided to set, removed from the distributions a growing part of the lightest isotopes when $Z$ is decreasing. The most neutron-rich Pt and Ir isotopes have not been measured because of missing settings of the magnetic fields during the experiment.

\subsection{Total cross sections}

\begin{table}[ht]
\begin{center}
\begin{tabular}{l  c  c  c}
  \hline
  Reaction               & $^{208}$Pb+p & $^{208}$Pb+p & $^{208}$Pb+d  \\
                         & (500A MeV)   & (1A GeV)     & (1A GeV) \\
  \hline
  \hline
  Spallation-evaporation (measured) & 1.44 (0.21)  & 1.68 (0.22)  & 1.91 (0.24)\\
  Total (measured)       & 1.67 (0.23)  & 1.84 (0.23)  & 2.08 (0.24) \\
  Total (calculated)     & 1.70         & 1.80         & 2.32 \\
  \hline
\end{tabular}
\caption{Spallation cross sections (in barns) for reactions $^{208}$Pb+p at 500A~MeV and 1A~GeV, and $^{208}$Pb+d at 1A~GeV. The measured spallation-evaporation and total cross sections (adding fission) are compared to a Glauber calculation performed using updated density distributions. Values in parenthesis are the total uncertainty of the measurements.}
\label{tab:total_cs}
\end{center}
\end{table}

We have estimated the total production cross-section of evaporation residues by summing all the measured residue cross sections, obtaining a value of (1.44$\pm$0.21)~b. Our measurement does not strictly cover all the range of the possible residues. However, the nuclides for which we have no measurement are mostly the lightest fragmentation products. Considering the steepness of the mass curve in this region (see figure~\ref{fig:cs_mass}), we can assume that the contribution of these nuclides is small, and probably much smaller than the error bars.

Adding the fission cross-section for this reaction~\cite{NPA_Bea} we estimated the total cross section to (1.67 $\pm$ 0.23) b.

This value is very close to the 1.70~b found by a Glauber-type calculation performed with updated density distributions. On table~\ref{tab:total_cs} we compare those values with the ones obtained in the reactions Pb+p and Pb+d at 1A~GeV. Although the slight decrease of the total cross section with respect to 1A~GeV measurements is in agreement with the expected trend, the 500A~MeV fission cross section is higher than previous measurements conducted at this energy, and also higher than the systematics of Prokofiev~\cite{Prokofiev}. This question has been discussed in detail in the corresponding paper~\cite{NPA_Bea}. Let us only underline that the agreement of a well-established model with our experimental total cross section comes in support of our measurement.

\subsection{Comparison to radiochemical measurements}

In recent years, a large number of measurements of spallation residues have been performed by the team of R. Michel. Of special relevance to our work is the measurement of residues of spallation of natural lead by protons at 550~MeV published by Gloris \etal~\cite{Gloris}. Cross sections with independent yields ({\it i.e.} nuclides that are not produced by $\beta$ decay) can be compared directly, while cross sections corresponding to accumulation of $\beta$-decaying nuclei require a summation of our data along the decay chain.

\begin{figure}[ht]
\begin{center}
\includegraphics*[width=0.9\textwidth]{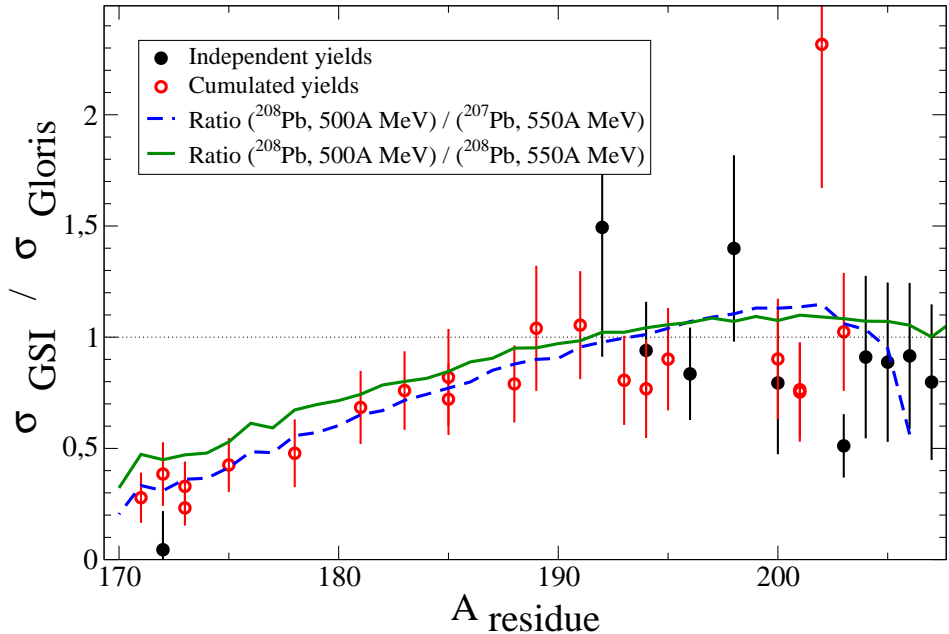}
\caption{Ratio between production cross-sections of residues measured in the reaction $^{208}$Pb+p at 500A~MeV (this work) and in the reaction p+$^{nat}$Pb at 550~MeV as a function of the mass of the residue for the isotopes measured in~\cite{Gloris}. Filled and empty circles represent nuclides with independent yields and cumulated yields, respectively. Calculations of the ratio between production cross sections at 550 and 500 MeV have been performed in two systems: with the same $^{208}$Pb target nuclei (continuous line) and with a different nucleus, $^{207}$Pb (dashed line), in order to study the effect of the use of natural lead in the Gloris experiment (see text).}
\label{fig:chemistry}
\end{center}
\end{figure}

The ratio between our data and those of Gloris \etal\ are presented in figure~\ref{fig:chemistry}. In the case of the cumulated yields, cross sections measured at GSI have been summed along the decay chain in order to be comparable with the radiochemical measurements. The agreement between the two data sets is overall fair for heavy residues, although a systematic shift of roughly 10\% may be guessed. Considering the error bars, all the measurements seem to be compatible, with the exception of $^{203}$Pb and $^{202}$Tl. As we have already underlined, our data points are very consistent with respect to one another. This makes such a large error in our measurement for these two nuclides rather unlikely, since it should have been clearly visible on our isotopic distributions.

For lighter nuclei the ratio decreases rapidly with increasing mass loss with respect to the projectile. This effect is the direct consequence of the differences between the measured systems: the 10\% higher energy strongly favors the production of lighter residues, up to a factor of 3 for mass losses around 35 nucleons.

This statement can be checked by using Monte-Carlo calculations. For this purpose we used the ISABEL~\cite{ISABEL} intranuclear cascade and the ABLA~\cite{ABLA} evaporation code. Although one of the purposes of these measurements is precisely to check the validity of those codes in the few hundreds of MeV region, we can assume that they are reliable enough if one only wants to calculate variations in a very limited energy and mass range, as it is the case here.

Results of the calculation of the ratio between isobaric cross sections for the two systems are also represented in figure~\ref{fig:chemistry}. A calculation that only takes into account the different incident energies offers a satisfactory reproduction of the decrease of the ratio for the light fragments. Replacing $^{208}$Pb by $^{207}$Pb (in order to mock the isotopic mixing of natural lead of which the targets of Gloris experiment were made) leads to a slight reduction of the calculated ratio for light nuclides, which improves the agreement with the data in this mass range. For heavy nuclides the calculations indicate that results at 500A~MeV should be larger than at 550A~MeV, which is not what we observed for most of the points. However, only 3 points are not compatible with the calculations when one considers the error bars. This leads us to conclude that, taking into account the differences between the systems measured in the Gloris experiment and our experiment, the agreement between these data sets is satisfactory.

\subsection{Mass spectra and comparison to previous GSI experiments}

\begin{figure}[ht]
\begin{center}
\includegraphics*[width=0.95\textwidth]{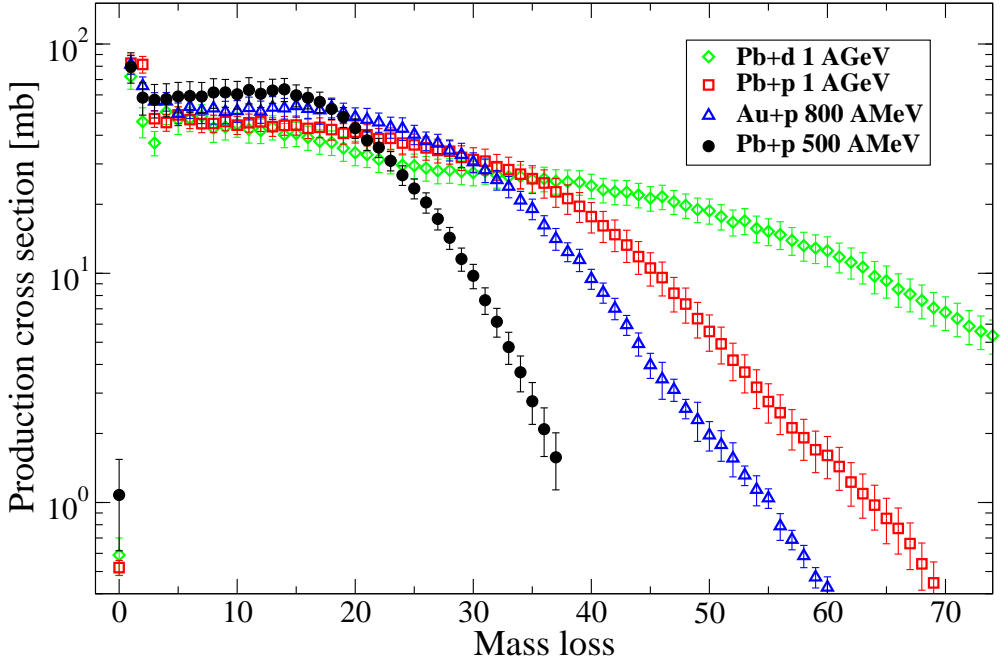}
\caption{Production cross-sections of the residues of the reaction Pb+p at 500A~MeV as a function of the mass loss with respect to the projectile (full circles). Data obtained in previously mentioned experiments are also represented: Au+p at 800A~MeV (triangles), Pb+p (squares) and Pb+d (diamonds) at 1A~GeV. The isolated points at $\Delta A=0$ correspond to a single nuclide, $^{208}$Bi.}
\label{fig:cs_mass}
\end{center}
\end{figure}

Figure~\ref{fig:cs_mass} presents the production cross sections, summed for all isobars, as a function of the mass loss with respect to the projectile. The data obtained from several experiments performed at the FRS are presented here: Pb+p at 500A~MeV, Pb+p at 1A~GeV~\cite{Pb_p}, Pb+d at 1A~GeV~\cite{Pb_d}, and Au+p at 800A~MeV~\cite{Au_Fanny}.

For small mass losses, each spectrum has a nearly constant value. In this mass range, the lower the incident energy, the higher the cross sections. With increasing mass losses, the cross sections start to decrease. Here, the lower the energy, the earlier and the steeper the fall. This is easily understood as the direct consequence of the exploration by the prefragment of all the possible range of excitation energy available in each system. In this respect, the measurement with deuterons gives insights about what would be obtained in a measurement conducted with protons at twice the energy. The shape of the mass-loss curve obtained from the measurement on gold is fully compatible with the tendencies observed for lead.

In the 500A~MeV experiment, a clear separation exists between the group of the evaporation products (which does not extend beyond mass losses of 40 mass units) and the group of the fission product (which starts around mass losses of 70~\cite{NPA_Bea}). This absence of mixing could also be demonstrated by studying the velocity spectra of the light fragments, which all exhibit a quasi-perfect Gaussian shape, while the presence of fission products would have introduced a characteristic double-bumped shape due to the forward-backward selection of the fission fragments by the FRS~\cite{Pb_p}.

\subsection{Charge pick-up}

\begin{figure}[ht]
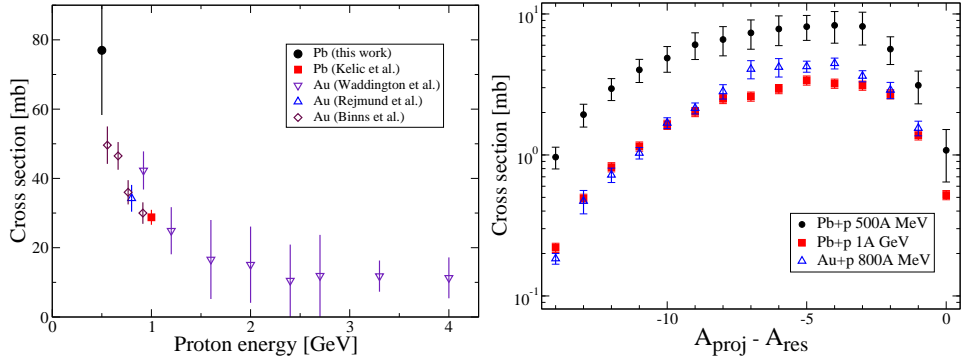

\begin{center}
\includegraphics*[width=0.45\textwidth]{charge_pickup.eps}
\includegraphics*[width=0.45\textwidth]{iso_pickup.eps}
\caption{Left figure: charge-pickup cross sections measured on lead (full symbols) and gold (empty symbols) as a function of the incident energy. The sum of the partial cross sections measured at the FRS (this work, full dots; Keli\'c \etal~\cite{Kelic}, full squares; Rejmund \etal~\cite{Au_Fanny}, upward triangles) is compared to elemental cross sections from Waddington \etal~\cite{Waddington} (downward triangles) and Binns \etal~\cite{Binns} (diamonds), which were both extracted from CH$_2$ and C measurements. Right figure: isotopic charge-pickup cross sections at 3 energies as a function of the mass loss with respect to the heavy partner of the reaction.}
\label{fig:pickup}
\end{center}
\end{figure}

An especially interesting result of this experiment is the measurement of the production cross section of 15 isotopes of Bi (see figure~\ref{fig:pickup}). Those nuclides are formed by charge-pickup reactions. In the energy range considered here, the capture of the incident proton is not initially possible, as the incident proton energy is well above the Fermi energy of the target nuclide. Therefore the formation of $^{209}$Bi is improbable, and the formation of $^{208}$Bi is only possible via, either the formation of a resonant state ($\Delta$ and pions), or a quasi-elastic collision between the incident proton and a neutron from the target nuclide, in which the neutrons leaves with an energy very close to the initial energy of the proton. The cross section for the charge-pickup is one of the few data that bring direct constraints for the intranuclear-cascade codes.

In the left part of figure~\ref{fig:pickup} we compare our measurement of the total charge-pickup cross section to previous measurements performed by our collaboration~\cite{Au_Fanny,Kelic}, Waddington \etal~\cite{Waddington} and Binns \etal~\cite{Binns}. For a qualitative discussion we do not need to discriminate between gold and lead as those nuclides are close to one another, both in atomic number and mass. Our measurement confirms the trend of a strong increase of the total cross section of the charge-pickup with decreasing energy.

This increase of the Bi production in Pb+p experiments concerns all Bi isotopes, as it can be seen in the right part of figure~\ref{fig:pickup}. The shapes of the 500A~MeV and 1A~GeV distributions are overall similar, but the overproduction at 500A~MeV increases slowly from a factor of 2 for the heaviest isotopes to a factor of 4 for the lightest. Problems in the separation of the ionic charge states~\cite{NIM_loa} prevented us to use the kinematic spectra to distinguish between the respective contribution of the $\Delta$ resonance and the quasi-elastic reactions in the formation of the heaviest Bi isotopes, as it was done by Keli\'c \etal~\cite{Kelic}.

The shape of the Hg spectrum (obtained in the Au+p at 800A~MeV measurement) is slightly different from the lead spectra. On one hand, for mass losses up to 7 mass units, the shape of the isotopic distribution is nearly identical to the Pb spectra, with values in-between the two Pb experiments, which is consistent with a smooth evolution as a function of the projectile energy. On the other hand the production of the lightest isotopes decreases faster than in the Pb+p experiments. This difference in shape can be explained by the lower Coulomb barrier and the shorter distance from the residue corridor~\cite{Dufour} for Hg nuclides with respect to Bi nuclides, which favor the emission of protons by the excited prefragments~\cite{Summerer_pickup}.

\subsection{Isobaric cross sections}

\begin{figure}[t]
\begin{center}
\includegraphics*[width=0.95\textwidth]{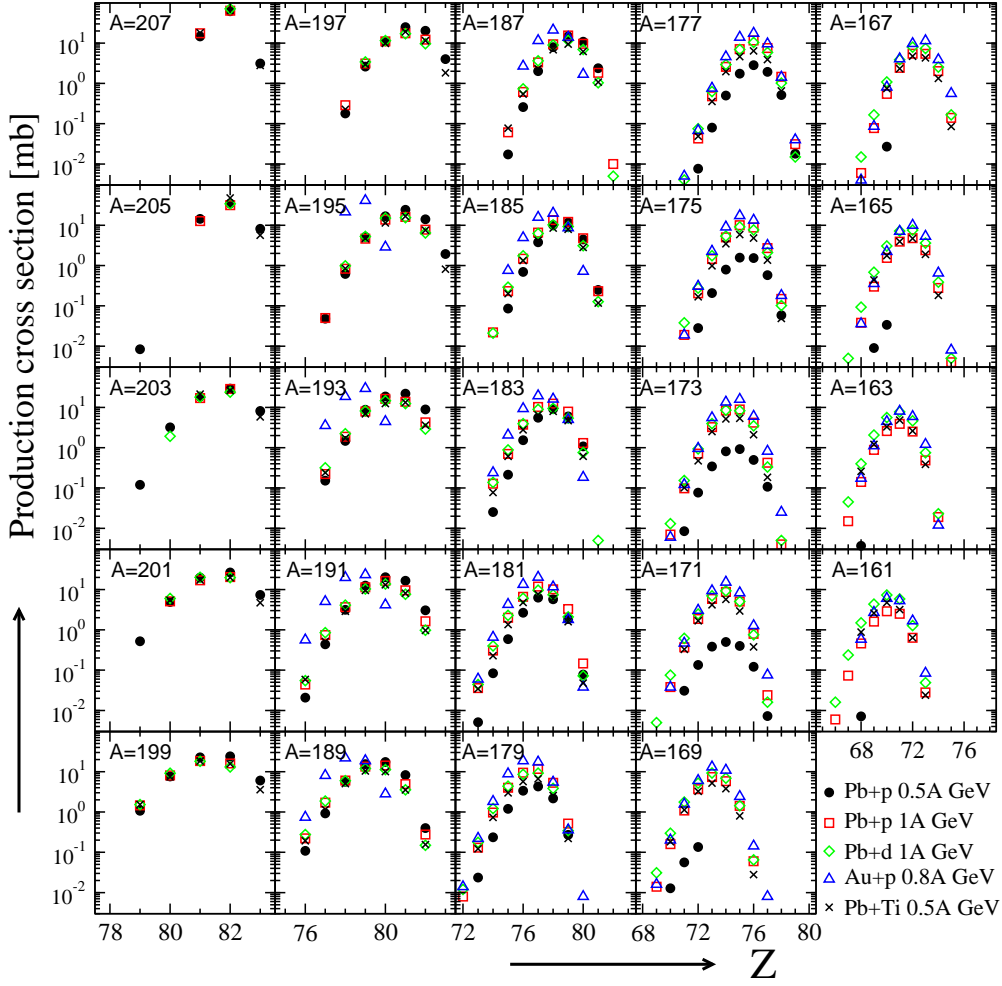}
\caption{Isobaric spectra of production cross-sections (in mb) in the reactions Pb+p and Pb+Ti at 500A~MeV (full circles and crosses, respectively). The data obtained in the experiments Au+p at 800A~MeV \cite{Au_Fanny}, Pb+p and Pb+d at 1A~GeV \cite{Pb_p,Pb_d} are also plotted (triangles, squares and diamonds, respectively). }
\label{fig:cs_isobaric}
\end{center}
\end{figure}

In figure~\ref{fig:cs_isobaric} the data from the same experiments as in previous sections are plotted as isobaric spectra.

\begin{figure}[t]
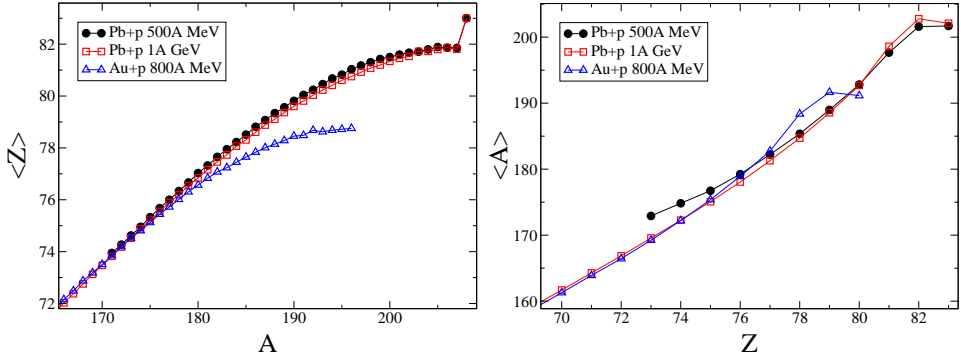

\begin{center}
\includegraphics*[width=0.45\textwidth]{mean_z.eps}
\includegraphics*[width=0.45\textwidth]{mean_a.eps}
\caption{Average atomic number as a function of the mass of the residue (left figure) and average mass of the residue as a function of the atomic number (right figure) in the reactions Pb+p at 500A~MeV (full circles, this work), Au+p at 800A~MeV (triangles, \cite{Au_Fanny}) and Pb+p at 1A~GeV (squares, \cite{Pb_p}). In the calculation of the average values, points have been weighted according to their cross section.}
\label{fig:cs_means}
\end{center}
\end{figure}

For heavy fragments, all distributions issued from reactions of Pb with p or d are very similar, both in shape and in magnitude. Low-energy reactions slightly dominate the cross sections for masses down to 185. With decreasing masses, the isobaric spectra behave in accordance with the mass distributions: the spectra for the highest-energy reaction scale down very slowly, whereas this scaling is steeper and steeper when one considers reactions at decreasing incident energy. However, for each isobar, the shapes of the different spectra remains extremely similar, as does its centroid (this can be seen in the left part of figure~\ref{fig:cs_means}).

Data from the reactions of Pb on the dummy target (which consists mainly of Ti in the target itself and Nb for the stripper foil placed after the target) at 500A~MeV have been added to the figure~\ref{fig:cs_isobaric} in order to illustrate the so-called limiting-fragmentation regime~\cite{limit_frag}. We observe no difference of shape or centroid between the spectra issued from the reaction on heavy ions and from the reaction on hydrogen isotopes.

The gold data offer an interesting point of comparison with the lead data. The gold fragments with mass close to 197 are associated with rather cold reactions and have therefore kept a $A/Z$ ratio very close to the initial system, while Pb fragments close to the same mass have lost roughly 10 nucleons, mostly neutrons because of the hindrance of charged-particle emission due to the Coulomb barrier. Therefore the gold and lead residue spectra are strongly shifted with respect to one another. This shift slowly vanishes with the increasing mass loss, which is easily understood as the slow move of the gold fragment distributions towards the residue corridor~\cite{Dufour}. This corridor is clearly visible on the left part of figure~\ref{fig:cs_means}: the barycenter of the isobaric distributions of the residues of all reactions converge on the same line.

This universal behavior, well known for reactions between heavy ions, is here demonstrated to be valid in a very broad energy range, even for fragments which are at the very end of the mass distribution. In other words, the isobaric distributions are independent of the incident energy in the system studied if properly renormalized. This is an experimental proof that the factorisation hypothesis is valid at energies as low as a few hundreds of MeV. This further strengthens the discussion regarding the agreement between EPAX and the data obtained from proton-induced reactions in systems in which fission does not play a major role (section~\ref{chap:secondary}).

Conversely, various projectile energies lead to variations of the center of the residues isotopic distributions, as it can be seen on the right part of the figure~\ref{fig:cs_means} as a deviation of the average mass value of the residues produced at 500A MeV. If this effect is washed out by the slow variations of the mass curve for higher energy reactions, it becomes noticeable at lower energies when the fall of the mass distribution becomes so steep that the production of the most neutron-deficient isotopes is strongly inhibited. Therefore, the reproduction of the isotopic and elemental cross-sections using a scaling factor between different systems is not appropriate at energies below the fragmentation limit.


\section{Conclusion}


The production cross sections and the momentum distributions have been measured for about 250 nuclei formed in the reaction of $^{208}$Pb on protons at 500A~MeV, covering most of the nuclides created down to a mass loss of 40 units with respect to the projectile, and with cross sections as low as 5~$\mu$b. The reaction products were identified in-flight in atomic number and mass-over-ionic-charge using the FRS spectrometer. The large proportion of non-fully stripped ions in the spectrometer was accounted for in detail, thus allowing to calculate the production cross section for each nuclide. The contribution of multiple reactions in the target to the residue production was carefully subtracted.

The production cross sections are in good agreement with previous radiochemical measurements. The isobaric distributions of the production cross sections are found to be very close to the ones measured at higher energies, thus extending the validity range of the factorisation hypothesis to energies of a few hundreds of MeV. The large variations observed on the isotopic cross sections can be nearly fully ascribed to the variations of the residue distributions with mass-loss at decreasing energy. Kinematical data are consistent with previous measurements.

The data obtained in this experiment, combined with previous measurements performed with the same technique (especially in the same system at 1A~GeV), constitute a set of information that is highly relevant for the development of reliable nuclear-reaction codes and, thus, the design of ADS.

\appendix


\section{Measured production cross sections}\label{chap:annexe_xs}


This appendix presents the spallation-evaporation production cross sections for the spallation of $^{208}$Pb by protons at 500A~MeV, expressed in mb. The value in the first parenthesis corresponds to the systematic error, and the value in the second parenthesis corresponds to the statistical error.

\fontsize{9pt}{11pt}
\selectfont

\begin{multicols}{3}
\setlength{\columnsep}{30pt}

\newcommand
\newelement{\begin{minipage}[t]{\columnwidth}{
\hbox{\raisebox{0.4em}{\rule{\columnwidth}{0.5pt}} }
\hspace{2pt}Z ~ A \ \  ~~~~~ $\sigma$ [mb] \ \ \ \ \ \ \\
\hbox{\raisebox{0.4em}{\rule{\columnwidth}{0.5pt}}}
}
\end{minipage}\\}

\begin{tabbing}

\newelement
83 \= 208 \= \  1.08 \= (0.44)(0.03) \\
83 \= 207 \= \  3.12 \= (0.82)(0.04) \\
83 \= 206 \= \  5.63 \= (1.26)(0.05) \\
83 \= 205 \= \  8.16 \= (2.14)(0.07) \\
83 \= 204 \= \  8.30 \= (2.10)(0.07) \\
83 \= 203 \= \  8.13 \= (1.63)(0.07) \\
83 \= 202 \= \  7.83 \= (1.87)(0.09) \\
83 \= 201 \= \  7.35 \= (1.73)(0.08) \\
83 \= 200 \= \  6.59 \= (1.54)(0.10) \\
83 \= 199 \= \  6.05 \= (1.30)(0.08) \\
83 \= 198 \= \  4.87 \= (1.01)(0.08) \\
83 \= 197 \= \  4.01 \= (0.75)(0.07) \\
83 \= 196 \= \  2.95 \= (0.52)(0.08) \\
83 \= 195 \= \  1.93 \= (0.36)(0.06) \\
83 \= 194 \= \  0.97 \= (0.17)(0.08) \\
\\
\newelement
82 \= 207 \= \  61.80 \= (9.05)(0.29) \\
82 \= 206 \= \  39.74 \= (6.02)(0.10) \\
82 \= 205 \= \  34.53 \= (5.18)(0.10) \\
82 \= 204 \= \  31.55 \= (4.74)(0.10) \\
82 \= 203 \= \  29.20 \= (4.39)(0.12) \\
82 \= 202 \= \  28.68 \= (4.24)(0.10) \\
82 \= 201 \= \  26.54 \= (3.94)(0.10) \\
82 \= 200 \= \  25.43 \= (3.69)(0.13) \\
82 \= 199 \= \  24.06 \= (3.40)(0.10) \\
82 \= 198 \= \  21.93 \= (3.05)(0.11) \\
82 \= 197 \= \  20.21 \= (2.72)(0.12) \\
82 \= 196 \= \  16.92 \= (2.26)(0.12) \\
82 \= 195 \= \  14.07 \= (1.78)(0.10) \\
82 \= 194 \= \  11.99 \= (1.45)(0.10) \\
82 \= 193 \= \   8.89 \= (1.03)(0.09) \\
82 \= 192 \= \   5.34 \= (0.61)(0.06) \\
82 \= 191 \= \   3.06 \= (0.34)(0.04) \\
82 \= 190 \= \   1.65 \= (0.19)(0.04) \\
82 \= 189 \= \   0.40 \= (0.05)(0.01) \\
82 \= 188 \= \   0.10 \= (0.01)(0.005) \\
\\
\newelement
81 \= 207 \= \  14.62 \= (1.72)(0.12) \\
81 \= 206 \= \  12.49 \= (1.32)(0.58) \\
81 \= 205 \= \  14.39 \= (1.89)(0.12) \\
81 \= 204 \= \  15.72 \= (1.98)(0.08) \\
81 \= 203 \= \  18.39 \= (2.23)(0.08) \\
81 \= 202 \= \  18.64 \= (2.28)(0.08) \\
81 \= 201 \= \  19.72 \= (2.41)(0.07) \\
81 \= 200 \= \  22.55 \= (2.65)(0.10) \\
81 \= 199 \= \  22.72 \= (2.70)(0.09) \\
81 \= 198 \= \  22.10 \= (2.67)(0.10) \\
81 \= 197 \= \  24.82 \= (2.86)(0.11) \\
81 \= 196 \= \  23.96 \= (2.75)(0.10) \\
81 \= 195 \= \  24.19 \= (2.70)(0.12) \\
81 \= 194 \= \  24.72 \= (2.64)(0.14) \\
81 \= 193 \= \  21.85 \= (2.29)(0.13) \\
81 \= 192 \= \  19.05 \= (1.94)(0.14) \\
81 \= 191 \= \  16.50 \= (1.62)(0.12) \\
81 \= 190 \= \  12.27 \= (1.18)(0.09) \\
81 \= 189 \= \  8.23 \= (0.78)(0.07) \\
81 \= 188 \= \  4.93 \= (0.47)(0.04) \\
81 \= 187 \= \  2.38 \= (0.23)(0.03) \\
81 \= 186 \= \  0.92 \= (0.09)(0.02) \\
81 \= 185 \= \  0.25 \= (0.03)(0.006) \\
81 \= 184 \= \  0.045 \= (0.006)(0.003) \\
\\
\newelement
80 \= 206 \= \   0.50 \= (0.04)(0.01) \\
80 \= 204 \= \   1.98 \= (0.19)(0.23) \\
80 \= 203 \= \   3.19 \= (0.26)(0.14) \\
80 \= 202 \= \   3.92 \= (0.37)(0.05) \\
80 \= 201 \= \   4.91 \= (0.45)(0.05) \\
80 \= 200 \= \   6.22 \= (0.73)(0.04) \\
80 \= 199 \= \   7.60 \= (0.69)(0.05) \\
80 \= 198 \= \   9.55 \= (1.07)(0.05) \\
80 \= 197 \= \  11.19 \= (1.24)(0.05) \\
80 \= 196 \= \  13.13 \= (1.44)(0.06) \\
80 \= 195 \= \  16.70 \= (1.71)(0.08) \\
80 \= 194 \= \  18.15 \= (1.84)(0.09) \\
80 \= 193 \= \  18.63 \= (1.88)(0.07) \\
80 \= 192 \= \  20.91 \= (2.02)(0.10) \\
80 \= 191 \= \  20.24 \= (1.93)(0.10) \\
80 \= 190 \= \  18.74 \= (1.77)(0.08) \\
80 \= 189 \= \  17.41 \= (1.60)(0.10) \\
80 \= 188 \= \  14.26 \= (1.30)(0.07) \\
80 \= 187 \= \  10.79 \= (0.98)(0.05) \\
80 \= 186 \= \   8.00 \= (0.72)(0.04) \\
80 \= 185 \= \   4.44 \= (0.41)(0.03) \\
80 \= 184 \= \   2.21 \= (0.21)(0.02) \\
80 \= 183 \= \   1.08 \= (0.10)(0.01) \\
80 \= 182 \= \   0.38 \= (0.04)(0.007) \\
80 \= 181 \= \   0.08 \= (0.01)(0.002) \\
80 \= 180 \= \   0.009 \= (0.002)(0.001) \\
\\
\newelement
79 \= 205 \= \   0.008 \= (0.006)(0.001) \\
79 \= 204 \= \   0.06 \= (0.02)(0.003) \\
79 \= 203 \= \   0.12 \= (0.03)(0.01) \\
79 \= 202 \= \   0.28 \= (0.06)(0.02) \\
79 \= 201 \= \   0.52 \= (0.09)(0.03) \\
79 \= 200 \= \   0.69 \= (0.10)(0.02) \\
79 \= 199 \= \   1.07 \= (0.10)(0.02) \\
79 \= 198 \= \   1.90 \= (0.22)(0.02) \\
79 \= 197 \= \   2.62 \= (0.30)(0.02) \\
79 \= 196 \= \   3.34 \= (0.39)(0.03) \\
79 \= 195 \= \   5.12 \= (0.52)(0.03) \\
79 \= 194 \= \   6.55 \= (0.65)(0.04) \\
79 \= 193 \= \   8.74 \= (0.84)(0.06) \\
79 \= 192 \= \  10.69 \= (1.00)(0.06) \\
79 \= 191 \= \  12.52 \= (1.17)(0.07) \\
79 \= 190 \= \  14.45 \= (1.17)(0.23) \\
79 \= 189 \= \  15.38 \= (1.39)(0.10) \\
79 \= 188 \= \  15.51 \= (1.39)(0.09) \\
79 \= 187 \= \  14.63 \= (1.31)(0.09) \\
79 \= 186 \= \  13.84 \= (1.23)(0.07) \\
79 \= 185 \= \  11.64 \= (1.04)(0.06) \\
79 \= 184 \= \   8.66 \= (0.79)(0.04) \\
79 \= 183 \= \   5.95 \= (0.55)(0.03) \\
79 \= 182 \= \   3.86 \= (0.37)(0.02) \\
79 \= 181 \= \   1.85 \= (0.19)(0.01) \\
79 \= 180 \= \   0.92 \= (0.10)(0.01) \\
79 \= 179 \= \   0.27 \= (0.03)(0.003) \\
79 \= 178 \= \   0.082 \= (0.012)(0.002) \\
79 \= 177 \= \   0.018 \= (0.003)(0.001) \\
\\
\newelement
78 \= 197 \= \   0.18 \= (0.03)(0.01) \\
78 \= 196 \= \   0.35 \= (0.04)(0.006) \\
78 \= 195 \= \   0.62 \= (0.07)(0.008) \\
78 \= 194 \= \   0.96 \= (0.10)(0.01) \\
78 \= 193 \= \   1.48 \= (0.15)(0.01) \\
78 \= 192 \= \   2.05 \= (0.21)(0.01) \\
78 \= 191 \= \   3.14 \= (0.30)(0.03) \\
78 \= 190 \= \   4.28 \= (0.40)(0.03) \\
78 \= 189 \= \   5.60 \= (0.52)(0.04) \\
78 \= 188 \= \   6.59 \= (0.62)(0.04) \\
78 \= 187 \= \   7.90 \= (0.73)(0.05) \\
78 \= 186 \= \   9.33 \= (0.84)(0.05) \\
78 \= 185 \= \  10.07 \= (0.91)(0.05) \\
78 \= 184 \= \  10.01 \= (0.91)(0.06) \\
78 \= 183 \= \   9.10 \= (0.84)(0.05) \\
78 \= 182 \= \   7.49 \= (0.71)(0.03) \\
78 \= 181 \= \   5.73 \= (0.57)(0.02) \\
78 \= 180 \= \   3.95 \= (0.41)(0.02) \\
78 \= 179 \= \   2.16 \= (0.25)(0.01) \\
78 \= 178 \= \   1.20 \= (0.15)(0.006) \\
78 \= 177 \= \   0.51 \= (0.07)(0.004) \\
78 \= 176 \= \   0.19 \= (0.03)(0.003) \\
78 \= 175 \= \   0.058 \= (0.011)(0.001) \\
\\
\newelement
77 \= 195 \= \   0.048 \= (0.004)(0.004) \\
77 \= 194 \= \   0.080 \= (0.007)(0.004) \\
77 \= 193 \= \   0.15 \= (0.01)(0.005) \\
77 \= 192 \= \   0.26 \= (0.02)(0.005) \\
77 \= 191 \= \   0.44 \= (0.04)(0.007) \\
77 \= 190 \= \   0.66 \= (0.05)(0.01) \\
77 \= 189 \= \   0.92 \= (0.09)(0.01) \\
77 \= 188 \= \   1.36 \= (0.14)(0.01) \\
77 \= 187 \= \   2.01 \= (0.20)(0.02) \\
77 \= 186 \= \   2.78 \= (0.27)(0.02) \\
77 \= 185 \= \   3.77 \= (0.36)(0.03) \\
77 \= 184 \= \   4.74 \= (0.44)(0.03) \\
77 \= 183 \= \   5.53 \= (0.52)(0.03) \\
77 \= 182 \= \   6.05 \= (0.57)(0.03) \\
77 \= 181 \= \   6.27 \= (0.60)(0.03) \\
77 \= 180 \= \   5.23 \= (0.54)(0.02) \\
77 \= 179 \= \   4.29 \= (0.47)(0.02) \\
77 \= 178 \= \   3.03 \= (0.37)(0.01) \\
77 \= 177 \= \   1.94 \= (0.27)(0.01) \\
77 \= 176 \= \   1.17 \= (0.18)(0.006) \\
77 \= 175 \= \   0.58 \= (0.10)(0.004) \\
77 \= 174 \= \   0.25 \= (0.05)(0.002) \\
77 \= 173 \= \   0.11 \= (0.02)(0.002) \\
77 \= 172 \= \   0.032 \= (0.008)(0.001) \\
77 \= 171 \= \   0.007 \= (0.002)(0.001) \\
\\
\newelement
76 \= 191 \= \   0.021 \= (0.004)(0.002) \\
76 \= 190 \= \   0.052 \= (0.006)(0.002) \\
76 \= 189 \= \   0.11 \= (0.01)(0.003) \\
76 \= 188 \= \   0.19 \= (0.02)(0.004) \\
76 \= 187 \= \   0.26 \= (0.03)(0.005) \\
76 \= 186 \= \   0.45 \= (0.05)(0.008) \\
76 \= 185 \= \   0.69 \= (0.07)(0.009) \\
76 \= 184 \= \   0.98 \= (0.10)(0.01) \\
76 \= 183 \= \   1.53 \= (0.15)(0.01) \\
76 \= 182 \= \   2.20 \= (0.21)(0.02) \\
76 \= 181 \= \   2.66 \= (0.26)(0.02) \\
76 \= 180 \= \   3.12 \= (0.32)(0.02) \\
76 \= 179 \= \   3.37 \= (0.36)(0.02) \\
76 \= 178 \= \   3.49 \= (0.38)(0.02) \\
76 \= 177 \= \   2.83 \= (0.35)(0.02) \\
76 \= 176 \= \   2.23 \= (0.31)(0.01) \\
76 \= 175 \= \   1.53 \= (0.25)(0.008) \\
76 \= 174 \= \   0.96 \= (0.18)(0.005) \\
76 \= 173 \= \   0.50 \= (0.12)(0.003) \\
76 \= 172 \= \   0.25 \= (0.07)(0.002) \\
76 \= 171 \= \   0.12 \= (0.04)(0.001) \\
\\
\newelement
75 \= 187 \= \   0.017 \= (0.002)(0.001) \\
75 \= 186 \= \   0.043 \= (0.005)(0.002) \\
75 \= 185 \= \   0.09 \= (0.01)(0.002) \\
75 \= 184 \= \   0.14 \= (0.01)(0.003) \\
75 \= 183 \= \   0.21 \= (0.02)(0.004) \\
75 \= 182 \= \   0.35 \= (0.04)(0.005) \\
75 \= 181 \= \   0.59 \= (0.06)(0.006) \\
75 \= 180 \= \   0.89 \= (0.09)(0.01) \\
75 \= 179 \= \   1.19 \= (0.13)(0.01) \\
75 \= 178 \= \   1.52 \= (0.17)(0.01) \\
75 \= 177 \= \   1.74 \= (0.20)(0.01) \\
75 \= 176 \= \   1.74 \= (0.22)(0.01) \\
75 \= 175 \= \   1.57 \= (0.23)(0.01) \\
75 \= 174 \= \   1.24 \= (0.22)(0.008) \\
75 \= 173 \= \   0.92 \= (0.20)(0.005) \\
75 \= 172 \= \   0.61 \= (0.17)(0.004) \\
75 \= 171 \= \   0.40 \= (0.13)(0.003) \\
\\
\newelement
74 \= 184 \= \   0.014 \= (0.001)(0.001) \\
74 \= 183 \= \   0.025 \= (0.002)(0.002) \\
74 \= 182 \= \   0.049 \= (0.004)(0.003) \\
74 \= 181 \= \   0.083 \= (0.008)(0.004) \\
74 \= 180 \= \   0.14 \= (0.01)(0.003) \\
74 \= 179 \= \   0.24 \= (0.02)(0.005) \\
74 \= 178 \= \   0.38 \= (0.04)(0.006) \\
74 \= 177 \= \   0.50 \= (0.06)(0.006) \\
74 \= 176 \= \   0.68 \= (0.08)(0.007) \\
74 \= 175 \= \   0.79 \= (0.10)(0.009) \\
74 \= 174 \= \   0.94 \= (0.13)(0.01) \\
74 \= 173 \= \   0.81 \= (0.14)(0.008) \\
74 \= 172 \= \   0.69 \= (0.15)(0.005) \\
74 \= 171 \= \   0.50 \= (0.15)(0.003) \\
\\
\newelement
73 \= 181 \= \   0.005 \= (0.001)(0.001) \\
73 \= 180 \= \   0.011 \= (0.001)(0.001) \\
73 \= 179 \= \   0.024 \= (0.003)(0.001) \\
73 \= 178 \= \   0.043 \= (0.005)(0.001) \\
73 \= 177 \= \   0.080 \= (0.009)(0.002) \\
73 \= 176 \= \   0.13 \= (0.015)(0.002) \\
73 \= 175 \= \   0.21 \= (0.024)(0.003) \\
73 \= 174 \= \   0.26 \= (0.035)(0.004) \\
73 \= 173 \= \   0.34 \= (0.05)(0.004) \\
73 \= 172 \= \   0.37 \= (0.07)(0.004) \\
73 \= 171 \= \   0.38 \= (0.08)(0.004) \\
73 \= 170 \= \   0.30 \= (0.09)(0.003) \\
\\
\newelement
72 \= 177 \= \   0.008 \= (0.001)(0.001) \\
72 \= 176 \= \   0.013 \= (0.002)(0.001) \\
72 \= 175 \= \   0.028 \= (0.003)(0.001) \\
72 \= 174 \= \   0.049 \= (0.006)(0.001) \\
72 \= 173 \= \   0.077 \= (0.01)(0.002) \\
72 \= 172 \= \   0.11 \= (0.02)(0.002) \\
72 \= 171 \= \   0.13 \= (0.02)(0.002) \\
72 \= 170 \= \   0.16 \= (0.03)(0.002) \\
72 \= 169 \= \   0.14 \= (0.04)(0.002) \\
\\
\\
\\
\newelement
71 \= 173 \= \   0.009 \= (0.001)(0.001) \\
71 \= 172 \= \   0.019 \= (0.002)(0.001) \\
71 \= 171 \= \   0.031 \= (0.004)(0.001) \\
71 \= 170 \= \   0.041 \= (0.007)(0.001) \\
71 \= 169 \= \   0.056 \= (0.011)(0.001) \\
71 \= 168 \= \   0.066 \= (0.012)(0.001) \\
\\
\\
\\
\\
\\
\newelement
70 \= 169 \= \   0.013 \= (0.002)(0.001) \\
70 \= 168 \= \   0.022 \= (0.003)(0.001) \\
70 \= 167 \= \   0.027 \= (0.006)(0.001) \\
70 \= 166 \= \   0.033 \= (0.010)(0.001) \\
70 \= 165 \= \   0.034 \= (0.014)(0.001) \\
\\
\newelement
69 \= 166 \= \   0.008 \= (0.001)(0.001) \\
69 \= 165 \= \   0.009 \= (0.003)(0.001) \\
69 \= 164 \= \   0.008 \= (0.004)(0.001) \\
\end{tabbing}
\end{multicols}
\normalsize


\section{Derivation of the contribution of the multiple reactions} \label{chap:sec_calc}


In this appendix we demonstrate the relation~\ref{eqn:secondary}. Let us consider a target of thickness $X$. At any point $x$ in this target, a nucleus $f$ can be produced by a nuclear reaction with a nucleus of the target, either from a beam nucleus or from a previously formed nucleus. On the other hand, an already formed $f$ nucleus can be destroyed. Formally, this can be written as:

\begin{equation} \label{eqn:sec}
\frac{dN_f(x)}{dx} = N_0(x) \sigma_{0\rightarrow f} - N_f(x) \sigma_f + \sum_{A_0 < A_i < A_f}^{} N_i \sigma_{i\rightarrow f}
\end{equation}

$\sigma_{a\rightarrow b}$ represents the cross section associated to the transformation of a nuclei $a$ in a nuclei $b$, $\sigma_n$ represents the total reaction cross section of a nuclide $n$, and $0$ labels correspond to the beam. In this calculation we assume that the variations of the cross sections related to the slowing down of the nuclei in the target are small enough to be neglected. The choice of the intermediate nuclei involved in the third term of the equation has been explained in section~\ref{chap:secondary}.

The number of nuclei in the beam is:

\begin{equation} \label{eqn:sec_beam}
N_0(x) = N_0(0) e^{-\sigma_0 x}
\end{equation}

The number of nuclei which are formed by a \textit{unique} reaction is obtained by solving the equation~\ref{eqn:sec}, ignoring the third term. This leads to:

\begin{equation} \label{eqn:sec_intermediate}
N_i(x) = \sigma_{0\rightarrow i} N_0(0) F(\sigma_i,\sigma_0, x)
\end{equation}

Here we used a function F defined as:

\begin{equation} \label{eqn:sec_defF}
F(\sigma_1,\sigma_2,x) = \frac{e^{-\sigma_1 x} - e^{-\sigma_2 x}}{\sigma_2 - \sigma_1}
\end{equation}

Replacing N$_0$ and $N_i$ by their expressions, equation~\ref{eqn:sec} is reduced to a first order differential equation, which is easily solved:

\begin{eqnarray}
\label{eqn:secrec}
N_f(x) & = & \sigma_{0\rightarrow f} N_0(0) F(\sigma_f,\sigma_0,x) \nonumber \\
       & + & \sum_{0 > i > f}^{} N_0(0) \; \sigma_{0\rightarrow i} \; \sigma_{i\rightarrow f} \;
          \left( F(\sigma_i,\sigma_f,x) - F(\sigma_0,\sigma_f,x) \right)
\end{eqnarray}

In the situation considered here, the products $\sigma x$ are always smaller than 0.1 (the total reaction probability in the target). We are then allowed to expand the exponentials, which leads to a simplified form of equation~\ref{eqn:secrec}:

\begin{equation}
\label{eqn:secrec2}
\sigma_{0\rightarrow f} = \frac{e^{\frac{\sigma_0+\sigma_f}{2}x}}{x} \left(
T_f(x) - \frac{x^2}{2} \sum_{0 > i > f}^{} \sigma_{0\rightarrow i} \; \sigma_{i\rightarrow f} \; e^{-\frac{\sigma_0+\sigma_i+\sigma_f}{3}x} \right)
\end{equation}

We have written the production rates of fragments $f$ as $T_f(x)$. This form is convenient because, knowing all the production cross sections from nuclide having a mass larger than the one from $f$ and the production rate of $f$, one can directly calculate the cross section for $f$. This system can now be solved by iterations over each isobar, starting from the mass of the projectile, providing one can calculate the various $\sigma$ terms.

One should note that the expression of equation~\ref{eqn:secrec2} is not only simpler than ~\ref{eqn:secrec}, but also more robust from a numerical point of view.



\vspace{1cm}

\begin{thebibliography}{9}

\bibitem{Bowmann} C. D. Bowman \etal\ - Nuclear Instruments and Methods A 320 (1992) 336
\bibitem{Rubbia} C. Rubbia \etal\ - Rapport CERN/AT/93-47 (1993)
\bibitem{NToF} C. Rubbia \etal\ - CERN/LHC/98-02(EET) and CERN/LHC/98-02(EET)-Add~1 (1998)
\bibitem{ESS} The ESS Project, vol. II, New Science and Technology for the 21st Century, ISBN 3-89336-302-5, http://neutron.neutron-eu.net/n\_ess/
\bibitem{ISOLDE} E. Kugler - Hyperfine Interactions 129 (2000) 23
\bibitem{EURISOL} The EURISOL report, available at www.ganil.fr/eurisol/Final\_Report.html
\bibitem{FRS} H. Geissel \etal\ - Nuclear Instruments and Methods B 70 (1992) 286
\bibitem{Au_Fanny} F. Rejmund \etal\ - Nuclear Physics A 683 (2001) 540
\bibitem{Au_Pepe} J. Benlliure \etal\ - Nuclear Physics A 683 (2001) 513
\bibitem{Wlazlo} W. Wlazlo \etal\ - Physics Review Letters 84 (2000) 5376
\bibitem{Pb_p} T. Enqvist \etal\ - Nuclear Physics A 686 (2001) 481
\bibitem{Pb_d} T. Enqvist \etal\ - Nuclear Physics A 703 (2002) 435
\bibitem{Taieb} J. Taieb \etal\ - Nuclear Physics A 724 (2003) 413-430
\bibitem{Monique} M. Bernas \etal\ - Nuclear Physics A 725 (2003) 213
\bibitem{U_Enrique} E. Casajeros - PhD thesis, Universitad de Santiago de Compostela (2003)
\bibitem{U_Jorge} J. Pereira - PhD thesis, Universitad de Santiago de Compostela (2004)
\bibitem{ISABEL} Y. Yariv, Z. Fraenkel - Physical Review C 20 (1979) 2227
\bibitem{INCL4} A. Boudard \etal\ - Physical Review C 66 (2002) 044615
\bibitem{BRIC} H. Duarte - Proc. of SARE-5 workshop on models and codes for spallation neutron sources (2000)
\bibitem{ABLA} A. Junghans \etal\ - Nuclear Physics A 629 (1998) 635
\bibitem{Dresner} L. W. Dresner - ORNL-TM-196 (1962)
\bibitem{Paolo} P. Napolitani et al. Physical Review C 70 (2004) 054607
\bibitem{NPA_Bea} B. Fern\'andez-Dom\'inguez \etal\ - Nuclear Physics A 747 (2005) 227
\bibitem{achromat} K.H. Schmidt \etal\ - Nuclear Instruments and Methods A 260 (1987) 287
\bibitem{NIM_loa} L. Audouin \etal\ - Nuclear Instruments and Methods A 548 (2005) 517

\bibitem{target} P. Chesny \etal\ - GSI Annual Report 97-1 (1997)
\bibitem{SEETRAM} B. Jurado \etal\ - Nuclear Instruments and Methods A 483 (2002), 603.
\bibitem{MUSIC} M. Pf\"utzner \etal\ - Nuclear Instruments and Methods B 86 (1994) 213
\bibitem{GLOBAL} C. Scheidenberger \etal\ - Nuclear Instruments and Methods B 142 (1998) 441
\bibitem{Karol} P. J. Karol - Physical Review C 11 (1975) 1203.

\bibitem{EPAX} K. S\"ummerer and B. Blank - Physical Review C 61 (2000) 034607
\bibitem{112Sn} A. Stolz \etal\ - Physical Review C 65 (2002) 064603
\bibitem{limit_frag} K. S\"ummerer \etal\ - Physical Review C 42 (1990) 2546

\bibitem{Morrissey} D.J. Morrissey - Physical Review C 39 (1989) 460
\bibitem{Goldhaber} M.L. Goldhaber - Physics Letters B 53 N 4 (1974) 306
\bibitem{fermi_mes} E. J. Moniz \etal\ - Physical Review Letters 26 (1971) 445
\bibitem{Hanelt} E. Hanelt et al. - Zeitschrift f\"ur Physik A 346 (1993) 43

\bibitem{Valentina} M.V. Ricciardi \etal\ - Nuclear Physics A 733 (2004) 299
\bibitem{Prokofiev} A.V. Prokofiev \etal\ - Nuclear Instruments and Methods A 463 (2001) 557
\bibitem{Gloris} M. Gloris \etal\ - Nuclear Instruments and Methods A 463 (2001) 593
\bibitem{Kelic} A. Keli\'c \etal\ - Physical Review C 70 (2004) 064608
\bibitem{Waddington} C.J. Waddington \etal\ - Physical Review C 61 (2000) 024910
\bibitem{Binns} W.R. Binns \etal\ - Physical Review C 39 (1989) 1785
\bibitem{Dufour} J.P. Dufour \etal\ - Nuclear Physics A 387 (1982) 157c
\bibitem{Summerer_pickup} K. S\"ummerer \etal\ - Physical Review C 52 (1995) 1106
\end{thebibliography}
\end{document}